\pdfoutput=1
\documentclass[options]{JHEP3}

\usepackage{amsmath,amssymb}

\ifpdf
  \usepackage[pdftex]{graphicx}
\else
  \usepackage[dvipdfmx]{graphicx}
\fi

\usepackage{ascmac}
\usepackage{ulem}
\usepackage{fancybox,calc}

\newcommand{\e}{\epsilon}
\newcommand{\oep}{\bar{\epsilon}}

\newcommand{\dl}{\delta}
\newcommand{\lam}{\lambda}
\newcommand{\olam}{\bar{\lambda}}
\newcommand{\si}{\sigma}
\newcommand{\g}{\gamma}

\newcommand{\Tr}{\text{Tr}}

\newcommand{\pa}{\partial}

\newcommand{\D}{\mathcal{D}}

\newcommand{\ophi}{\overline{\phi}}
\newcommand{\opsi}{\overline{\psi}}
\newcommand{\oF}{\overline{F}}

\newcommand{\na}{\nabla}

\title{Symmetry breaking caused by large $\mathcal{R}$-charge}
\author{Akinori Tanaka, Akio Tomiya and Takuya Shimotani\\ Department of Physics, Graduate School of Science, Osaka University,
 Toyonaka, Osaka 560-0043, Japan
 }

\preprint{OU-HET 814}

\abstract{
We discuss the gauge symmetry breaking via the Hosotani mechanism by using exact results on supersymmetric gauge theories based on the localization method.
We use the theories on $S^2 \times S^1$ Euclidean space, and study how the effective potential for the Wilson line phase varies by running an imaginary chemical potential.
In order to break the symmetry, we find that large $\mathcal{R}$-charge is necessary.
With such large $\mathcal{R}$-charge, we study the phase structure of the theory.
In addition, we observed that a finite size effect on our curved space when we take $\mathcal{R}$-charge is not so large.
}
\begin{document}
\renewcommand{\include}[1]{}
\renewcommand\documentclass[2][]{}

\section{Introduction}
SUSY gauge theories allow us to see various non-perturbative aspects of the quantum field theory which are not accessible with the conventional perturbation method.
We have well defined supersymmetry even on certain curved spacetimes in recent years,
and theories on such curved spacetimes have different characters compared with the usual theories on the flat spacetime.

Gauge symmetry is introduced as a origin of forces in quantum field theory.
The standard model has $\mathop{\rm SU}(2)\times \mathop{\rm U}(1)$ gauge symmetry.
This symmetry is broken to $\mathop{\rm U}(1)_{\text{EM}}$ phase by the Higgs mechanism.
A Higgs boson has been discovered by LHC experiments.
We are going to the next stage of Higgs search.
Although the standard model is successful theory, it does not contain the dark matter and it has naturalness problem.
Therefore we need to consider the beyond the standard model.

One of the attractive models with new physics at TeV scale is the gauge-Higgs unification scenario\cite{Hosotani:1983xw, Hosotani:1988bm, Davies:1987ei, Davies:1988wt, Hatanaka:1998yp, Burdman:2002se, Csaki:2002ur, Lim:2013eha}.
In the gauge-Higgs unification scenario, the gauge symmetry is broken by the Wilson line phase
 which comes from non-simply connected structure of compactified extra dimensional space.
In this context, we can regard a component of gauge field along the extra dimension as a 4 dimensional Higgs field.
This gauge field along the extra dimension could have a vacuum expectation value.
This value comes from the Wilson line phase because it cannot be gauged away.
We can get the effective potential for the Wilson line phase.
Surprisingly, although higher dimensional theories are non-renormalizable, this effective potential and Higgs mass are finite at 1 loop level.
Of course the gauge symmetry is never broken at the tree level because the Wilson line phase is a solution of the equations of motion which are invariant under the gauge transformation.
However, the gauge symmetry is broken spontaneously at the loop level.
This is called \textit{the Hosotani mechanism} which has been studied as a electroweak symmetry breaking mechanism\cite{Agashe:2004rs, Medina:2007hz, Hosotani:2008tx, Serone:2009kf, Hosotani:2009qf, Funatsu:2014fda, Maru:2013bja}.

However, there is an unsatisfactory point with Hosotani mechanism :
this mechanism for non-Abelian gauge theory has not been established by all order or non-perturvative way.
In order to overcome this problem, it is necessary to study it with \textit{the non-perturvative method}.
One naive way is to use the lattice gauge theory.
The mechanism have been studied already by lattice gauge theories\cite{Kashiwa:2013rmg, Cossu:2013ora, Cossu:2013nla, Kashiwa:2013rqa, Irges:2013rya, Itou:2014iya}.
For example, in \cite{Cossu:2013ora, Cossu:2013nla}, they studied the $\mathop{\rm SU}(3)$ gauge symmetry in the 3+1 dimensional flat spacetime.
In their analysis, SU(3) gauge theory with adjoint fermions has 4 phases, \textit{confined phase, deconfined phase, split phase} and \textit{reconfined phase}.
By changing the mass of the adjoint fermions, these distinct phases emerge in a certain order.
In terms of the Hosotani mechanism, they show that these phases correspond to $\mathop{\rm SU}(3)$, $\mathop{\rm SU}(3)$, $\mathop{\rm SU}(2)\times\mathop{\rm U}(1)$ and $\mathop{\rm U}(1)\times\mathop{\rm U}(1)$ global symmetries respectively.
In addition to the adjoint fermions, they also considered the fundamental fermions, and checked \textit{the Rogerge-Weiss (RW) transition}\cite{Roberge:1986mm}.

On the other hand, in these days, an exact way to perform the path integral so-called localization method have been developed \cite{Pestun:2007rz, Kapustin:2009kz, Benini:2012ui, Doroud:2012xw, Kallen:2012va}.
A novel point for these recent developments is defining SUSY gauge theories on a compact manifold in order to regularize IR divergence naturally.
For example, we can choose $M \times S^1$ as such a compact space.
It turns out to be possible to construct supersymmetry on $M \times S^1$ for a certain $M$, and the exact results are known as so-called (superconformal) index \cite{Kim:2009wb, Gang:2009wy, Imamura:2011su, Romelsberger:2005eg, Kim:2013nva, TanakaMori} which have fine informations about the BPS spectra for the theories.
If we consider SUSY gauge theories by taking 
\begin{align}
M= \text{a flat spacetime}, \notag 
\end{align}
the effective potential for the Wilson line phase turns to be totally flat because the fermionic contribution cancels the corresponding bosonic contribution. As a result, gauge symmetry is unbroken.
However, we define SUSY gauge theories by taking 
\begin{align}
M= \text{a curved spacetime}, \notag 
\end{align}
in this case, the boson only couples with the background scalar curvature.
This makes difference between bosons and fermions.
This fact may suggest a possibility towards a non-trivial and non-perturbative analysis of Hosotani mechanism based on SUSY gauge theories on a curved $M \times S^1$.
As a first step to move on more realistic studies, 
we analyze gauge theory on $S^2\times S^1$ in this paper for simplicity.

\paragraph{A formal argument}
In usual argument, a minimum of the effective potential is selected because of the large volume in the following sense.
Suppose we have a partition function as
\begin{align}
Z
=
\sum_{v \in \text{vacua}}
e^{- \text{Vol} \cdot V_{\text{eff}} ( v )}
.
\label{Z1}
\end{align}
When we take Vol$\to \infty$, the steepest decent $v_0$ will dominate $Z$.
It means the vacuum which satisfies
\begin{align}
&V_{\text{eff}}' (v_0) =0,
\label{var1}
\\
&V_{\text{eff}}'' (v_0) >0,
\label{var2}
\end{align}
is selected automatically as a true vacuum.
In this paper, we use $\mathcal{N} =2$ superconformal field theories.
Intuitively, we have no volume dependence with these theories because of the conformal symmetry.
However, as we noted above, once we turn on the matter fields into the theory on a certain curved space,
the matter couples with the scalar curvature R via \textit{$\mathcal{R}$-charge} $\Delta_{\Phi}$.
Our discussion on the symmetry breaking is based on \textit{large $\mathcal{R}$-charge limit} $\Delta_{\Phi} \to \infty$.
Schematically, in our case, the partition function takes the following form
\begin{align}
Z
=
\sum_{v \in \text{vacua}}
e^{- \Delta_{\Phi} \cdot V_{\text{eff}} ( v )}
.
\label{Z2}
\end{align}
Through the same argument presented above, when we take $\Delta_{\Phi} \to \infty$, a vacuum corresponds to \eqref{var1}, \eqref{var2} is selected.
Actually, large $\mathcal{R}$-charge limit is same as the thermodynamic limit. We will argue this issue in Section 5.

This paper is organized as follows.
In Section 2, we summarize results of exact calculation for super Yang-Mills (SYM) theory with two matters on $S^2 \times S^1$.
And we discuss large $\mathcal{R}$-charge limit which causes the symmetry breaking.
In Section 3, we investigate \textit{the finite size effects} via the effective potential with the small $\mathcal{R}$-charge.
In Section 4, we show phase structures of the broken vacua at large $\mathcal{R}$-charge limit.
Section 5 contains results and comments on our method.
In Appendix A we summarize our localization calculous.
\section{Preliminaries}
\subsection{Basic concepts}
\paragraph{Our spacetime}
What we want to discuss is the symmetry breaking via the Wilson line phase along $S^1$ extra dimension.
In addition, we want to take nontrivial $\mathcal{R}$-charge contribution into account via the coupling with the scalar curvature R.
As a simplest model which satisfies these conditions, we choose a gauge theory on $S^2 \times S^1$(Figure \ref{S2S1}).
We do not consider Minkowskian theories but Euclidean ones throughout this paper.
\begin{figure}[t]
  \begin{center}
   \includegraphics[width=70mm]{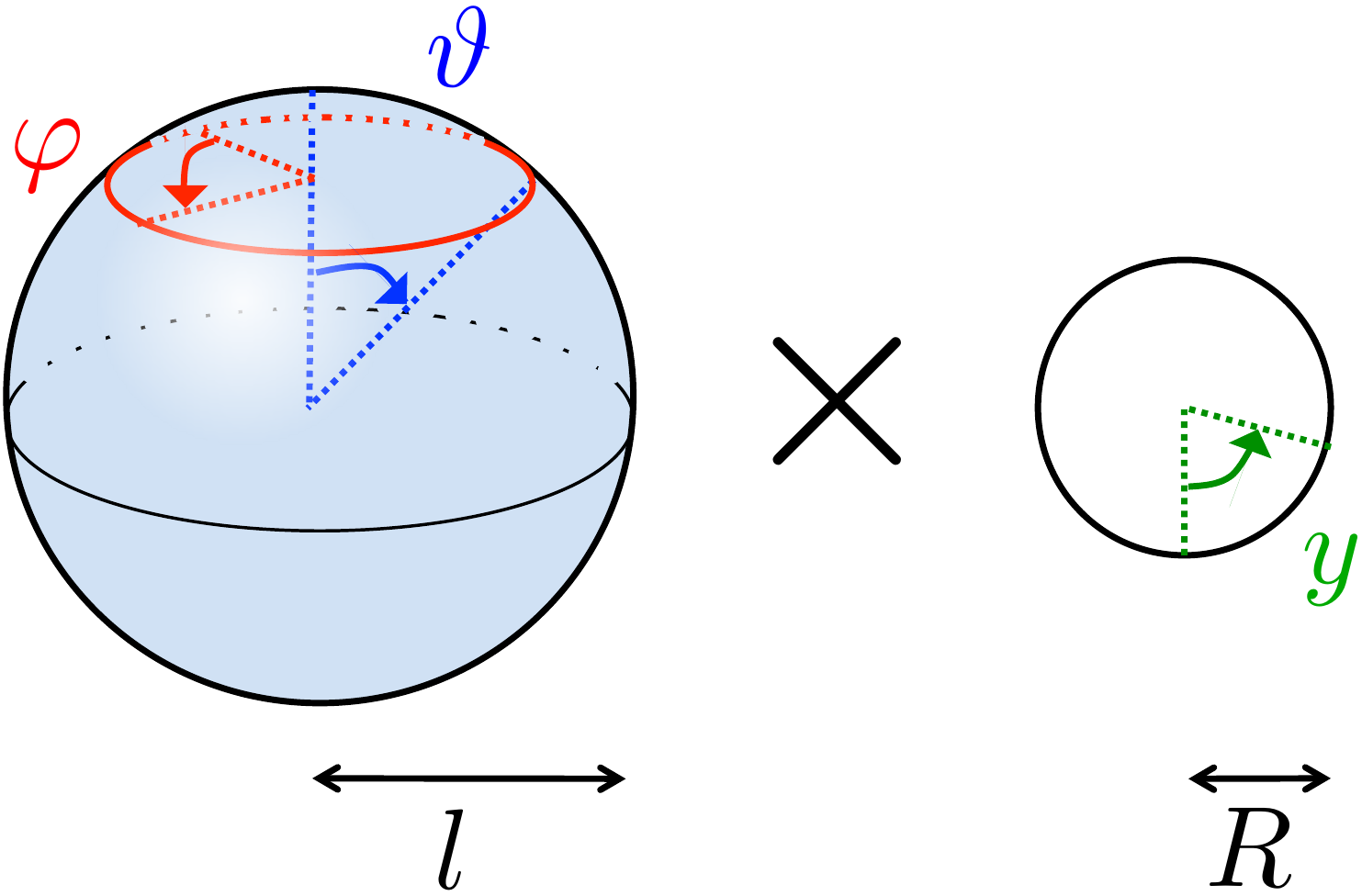}
  \end{center}
  \caption{$S^2 \times S^1$, $l$ is the radius of $S^2$, and $R$ is the radius of $S^1$.}
  \label{S2S1}
 \end{figure}
Therefore, we take the following metric and coordinates of the $S^2 \times S^1$,
\begin{align}
&
ds^2 = l^2 (d \vartheta^2 + \sin ^2 \vartheta d \varphi^2) + d y^2,
\qquad
\vartheta \in[ 0 , \pi],
\quad
\varphi \in [0 , 2 \pi],
\quad
y \in [ 0 , 2 \pi R].
\label{met}
\end{align}
As reported in \cite{Kim:2009wb, Gang:2009wy,  Imamura:2011su},
even on such a curved space, we can construct supersymmetric field theories\footnote{See more details in  Appendix A.}.
Note that we have 3 distinct ``R" s,
\begin{center}
$\mathcal{R}$ - charge  : curly $\mathcal{R}$, \hspace{4mm}
$S^1$- radius  :  Itaric \textit{R}, \hspace{4mm}
scalar curvature  :  normal R.
\end{center}
\paragraph{Possible fields and theories}
We discuss later the nontrivial phases of SUSY gauge theory on $S^2 \times S^1$.
In order to clarify what we have done, we give here a lightning review of $\mathcal{N}=2$ off-shell supersymmetry on $S^2 \times S^1$.
We can construct two distinct irreducible field representations of $\mathcal{N}=2$ off-shell supersymmetry,
\begin{align}
\text{vector multiplet : }
&V=
(A_\mu , \sigma , \bar{\lambda}, \lambda, D)
\in \text{Ad},
\label{vec}\\
\text{matter multiplet : }
&
\Phi=
 (\phi , \psi , F)
 \in \text{Rep}
,
\end{align}
where the $\sigma, D, F$ are scalar fields, $\bar{\lambda}, \lambda, \psi$ are spinors and $A_\mu$ is a gauge field.
In the flat case, one can get these supermultiplets in 3 dimensional space by the dimensional reduction from 4 dimensional $\mathcal{N}=1$ vector and matter multiplets respectively.
By using these off-shell component fields, we can construct the following supersymmetric Lagrangians\footnote{We can also take the supersymmetric \textit{Chern-Simons} (CS) term. However it will cause a sign problem. Therefore, we discard the CS term in this paper for simplicity.},
\begin{align}
\mathcal{L}_{\text{SYM}}
=&
\Tr
\Big(
 \frac{1}{2}    F_{\mu \nu}  F^{\mu \nu}
 + D^2
 +\D_\mu \si  \D^\mu \si
 + \frac{1}{l}  \e^{3 \rho \si} \si F_{\rho \si} 
 + \frac{ \si^2 }{l^2} 
+ i \olam \g^\mu \D_\mu \lam
- i \olam [ \lam , \si]
- \frac{i}{2l} \olam \g_3 \lam
\Big),
\label{SYMl}
\\
\mathcal{L}_{\Phi}
=&
-i  ( \opsi \g^\mu \D_\mu\psi)
 +i ( \opsi \si  \psi) 
- i \ophi  (\olam \psi)
 - \frac{i(2\Delta_{\Phi} -1)}{2l} ( \opsi \g_3 \psi)  
  +\oF F
\notag \\ &
+ i (  \opsi \lam)  \phi 
+    \D_\mu \ophi \D^\mu \phi  
+   \ophi \si^2 \phi  
 + i  \ophi D  \phi  \
 - \frac{2\Delta_{\Phi} -1}{l}    \ophi \D_3 \phi  
- \frac{\Delta_\Phi(2\Delta_\Phi -1)}{2l^2}  \ophi \phi 
+ \frac{\Delta_\Phi}{4}\text{R} \ophi \phi,
\label{Smat}
\end{align}
where R is the scalar curvature calculated from \eqref{met} and $\Delta_\Phi$ is the $\mathcal{R}$-charge of the matter multiplet $\Phi$.
We can take arbitrary $\Delta_\Phi$ without breaking supersymmetry.
In addition, we define the covariant derivative $\D_\mu$ as
\begin{align}
\D_\mu = \na_\mu - i A_\mu,
\label{covd}
\end{align}
where $\na_\mu$ is the covariant derivative with respect to the spin connection:
\begin{align}
\na_\mu (\text{scalar}) = \pa_\mu (\text{scalar}),
\qquad
\na_\mu (\text{spinor}) = (\pa_\mu - \frac{1}{4} \omega{_\mu}{^{ab}} \g_{ab} ) (\text{spinor}).
\end{align}

\subsection{Our model and the vacua}
\paragraph{Field contents}
We consider $\mathop{\rm SU}(3)$ gauge theories which have been investigated in the context of the symmetry breaking via the Wilson line phase recently in \cite{Cossu:2013ora, Itou:2014iya} with the lattice gauge theory.
Our model is constructed by
\begin{align}
&\text{1 vector : }
V,
\notag \\
&\text{2 matters : }
\left\{ \begin{array}{ll}
\Phi_1
&
\text{represented by }
+\rho
\\
\Phi_2
&
\text{represented by }
-\rho
\end{array} \right. .
\end{align}
\paragraph{The $\mathcal{R}$-charge and chemical potential for matters}
In addition, in order to simplify the exact calculations by the supersymmetric localization method, we assign identical $\mathcal{R}$-charges $\Delta$ with these matters:
\begin{align}
\Delta_{\Phi_1} =
\Delta_{\Phi_2} =
 \Delta,
\end{align}
and turn on opposite imaginary chemical potentials 
\begin{align}
\mu_{\Phi_1}
=
-\mu_{\Phi_2}
=
i \alpha
.
\end{align}
The simplifications caused by this choice of quantities will be explained in Appendix A.
For later use, we comment on the boundary conditions of the component fields in the matter multiplets.
We have many fields which satisfy the boundary conditions \eqref{mbc1}-\eqref{mbc6}.
For example, the scalars $\phi_1, \ophi_1, \phi_2, \ophi_2$ satisfy
\begin{align}
&
\phi_1(\vartheta,\varphi , y + 2 \pi R)
=
e^{- \Delta \frac{\pi R}{l}}e^{ + i \alpha}
\phi_1(\vartheta,\varphi , y ),
\label{bcphi1}
\\
&\ophi_1(\vartheta,\varphi , y + 2 \pi R)
=
e^{+ \Delta \frac{\pi R}{l}}e^{ - i \alpha}
\ophi_1(\vartheta,\varphi , y ),
\label{bcophi1}
\\
&
\phi_2(\vartheta,\varphi , y + 2 \pi R)
=
e^{- \Delta \frac{\pi R}{l}}e^{ - i \alpha}
\phi_2(\vartheta,\varphi , y ),
\label{bcphi2}
\\
&\ophi_2(\vartheta,\varphi , y + 2 \pi R)
=
e^{+ \Delta \frac{\pi R}{l}}e^{ + i \alpha}
\ophi_2(\vartheta,\varphi , y ).
\label{bcophi2}
\end{align}
Note that the facters $e^{\pm \Delta \frac{\pi R}{l}}$ are necessary in order to maintain the supersymmetry.
\paragraph{Our Lagrangian and the vacua}
We have introduced SYM Lagrangian $\mathcal{L}_{\text{SYM}}$ in \eqref{SYMl} and matter Lagrangian $\mathcal{L}_\Phi$ in \eqref{Smat}.
Throughout this paper, we consider the following Lagrangian on $S^2 \times S^1$ :
\begin{align}
\mathcal{L}
=
\mathcal{L}_{\text{SYM}}
+
\mathcal{L}_{\Phi_1}
+
\mathcal{L}_{\Phi_2}
.
\end{align}
This Lagrangian gives the following vacua \cite{Kim:2009wb, Gang:2009wy, Imamura:2011su}, in other words, \textit{the locus} :
\begin{align}
&A= m A_{\text{mon}} + \frac{\theta}{2 \pi R} dy
,
\quad
\si = - \frac{m}{2l},
\notag
\\
&
m = \text{diag} (m_1, m_2, -m_1-m_2),
\notag
\quad
\theta= \text{diag} (\theta_1, \theta_2, -\theta_1-\theta_2),
\end{align}
other fields are zero.
$A_{\text{mon}}$ is the Dirac monopole configration:
\begin{align}
A_{\text{mon}} 
=
\frac{1}{2} (\kappa - \cos \vartheta) d \varphi
,
\quad
\kappa
=
\left\{ \begin{array}{ll}
+1& \quad \vartheta \in [0, \frac{\pi}{2}] \\
-1& \quad \vartheta \in [\frac{\pi}{2} , \pi]\\
\end{array} \right. .
\end{align}
The $m$ is so-called GNO charge \cite{Goddard:1976qe},
and $\theta$ represent the Wilson line phase.
\subsection{An exact result, large $\mathcal{R}$-charge limit and the symmetry breaking}
Via so-called localization method \cite{Kim:2009wb, Gang:2009wy, Imamura:2011su}, we can calculate the path integral on the $S^2\times S^1$ exactly, and get the result in the form of a summation over the locus:
\begin{align}
&\int \mathcal{D} V \mathcal{D} \Phi_1 \mathcal{D} \overline{\Phi}_1 \mathcal{D} \Phi_2 \mathcal{D} \overline{\Phi}_2
\ e^{- \int d^3 x \sqrt{g} \mathcal{L}}
&=
\sum_{m_1,m_2 = -\infty}^\infty
\frac{1}{\text{sym}}
\int_{-\pi}^\pi
\frac{d \theta_1}{2 \pi R}
\frac{d \theta_2}{2 \pi R} \
\mathcal{Z}_{\text{1-loop}}^{vec(reg)} 
\times
\mathcal{Z}_{\text{1-loop}}^{mat1,2(reg)} 
.
\label{index}
\end{align}
The `sym' represents symmetric factors for the configurations of $m_1,m_2$.
The summation $\sum_{m_1,m_2}$ comes from the GNO monopoles' quantization condition on $S^2$ which can be regarded as one of \textit{the finite size effects}.
We discuss details of the finite size effects later.
The integral $\int d \theta_1 d \theta_2$ is caused by the Wilson line phase.
The domain $[-\pi , \pi]$ is a consequence of the gauge symmetry of $\mathcal{L}$.
Let us turn to the integrands.
The first one is the contribution from the vector multiplet :
\begin{align}
&\mathcal{Z}_{\text{1-loop}} ^{vec(reg)} 
=
\prod_{i > j} 
\Big| 2
\sin
\Big( 
\frac{\theta_i - \theta_j}{2} + i \frac{\pi R}{l} \frac{m_i - m_j}{2}
 \Big)
 \Big|^2.
\label{2vecl3}
\end{align}
The second one is the contribution from the two matter multiplets :
\begin{align}
&\mathcal{Z}_{\text{1-loop}} ^{mat1,2(reg)}
=
\prod_{\rho \in R} 
\prod_{J = 1- \frac{\Delta}{2}}^{\frac{\Delta}{2}-1}
\Big| 2
\sin 
\Big( \frac{\rho(\theta) -\alpha}{2} +i \frac{\pi R}{l} \big(  |\frac{ \rho(m)}{2} | + J \big)
\Big) 
\Big|^2
,
\label{2matl5}
\end{align}
where we have assumed $\Delta -1 \in \mathbb{N}$.
Of course there is no $\Delta$ dependence on the contribution from the vector multiplet $\mathcal{Z}_{\text{1-loop}} ^{vec(reg)}$, 
but the contribution from the two matter multiplets $\mathcal{Z}_{\text{1-loop}} ^{mat1,2(reg)}$.
We can rewrite this $\mathcal{Z}_{\text{1-loop}} ^{mat1,2(reg)}$ into a more useful form
\begin{align}
\mathcal{Z}_{\text{1-loop}} ^{mat1,2(reg)}
&=
\exp{\Big(
- \Delta
V_{\text{eff}} (\theta,m)
\Big)},
\end{align}
where
\begin{align}
V_{\text{eff}} (\theta,m)
=
-
\sum_{\rho \in R}  
2\text{Re}
 \sum_{J = 1- \frac{\Delta}{2}}^{\frac{\Delta}{2}-1}
\frac{1}{\Delta}
 \log 2
\sin 
\Big( \frac{\rho(\theta) -\alpha}{2} +i \frac{\pi R}{l} \big(  |\frac{ \rho(m)}{2} | + J \big) \Big)
.
\label{eff}
\end{align}
Roughly speaking,
$V_{\text{eff}} \sim \sum_{J} \frac{1}{\Delta} \sim \Delta \frac{1}{\Delta} =1$.
Therefore, this definition is meaningful for any integer $\Delta$.
\paragraph{Large $\mathcal{R}$-charge limit}
In order to discuss the symmetry breaking, we take $\Delta \to \infty$ as we noted in the introduction.
However there is one problem.
See boundary conditions \eqref{bcphi1} - \eqref{bcophi2}.
These conditions include the following factor
\begin{align}
\Delta \frac{\pi R}{l} =: c.
\label{c}
\end{align}
The naive $\Delta \to \infty$ limit defines pathological behaviors for $\Phi_1, \Phi_2$
because $c \to \infty$.
Therefore, we have to take $\Delta \to \infty$ with fixing $c$ in order to avoid such ill defined $S^1$ boundary conditions for $\Phi_1, \Phi_2$.
This means, we have to take $l \to \infty$ together. It corresponds to the large volume limit.
Note that $R \to + 0$ is different from $l \to \infty$ because of the presence of $R$ in \eqref{index}.
Once we take $\Delta \to \infty$,
we can replace $\sum_J \frac{1}{\Delta}$ to the corresponding integral over $- 1/2$ to $1/2$ 
 in the sense of Riemann sum:
\begin{align}
\sum_{J = 1- \frac{\Delta}{2}}^{\frac{\Delta}{2}-1} \frac{1}{\Delta}
=
\sum_{J = 1- \frac{\Delta}{2}}^{\frac{\Delta}{2}-1} \frac{\dl J}{\Delta}
\to
\int _{-1/2}^{+1/2}
d j,
\end{align}
where we define a continuous parameter $j$ as
\begin{align}
j : = \frac{J}{\Delta}.
\end{align}
In addition, we can simplify
\begin{align}
\frac{\rho(\theta) -\alpha}{2} +i c\big(  |\frac{ \rho(m)}{2 \Delta} | + j \big)
\quad \to \quad \frac{\rho(\theta) -\alpha}{2} +i c  j ,
\label{zlim}
\end{align}
by using \eqref{c} and $\Delta \to \infty$.
In \eqref{zlim}, the $m$ dependence vanishes.
This is natural because $m$ dependence can be regarded as the finite size effect of $S^2$ which comes from the nontrivial Dirac monopole configuration.
In summary, our effective potential for the Wilson line phase is constructed by
\begin{align}
V_{\text{eff}} (\theta,m)
\to
\mathcal{V}_{\text{eff}} (\theta)
&=
-
\sum_{\rho \in R} 
2 \text{Re}
 \int_{-1/2}^{+1/2}
dj
\
\log 
\sin \big( \frac{\rho(\theta) -\alpha }{2} + i c j  )
\notag \\
&=
-
\sum_{\rho \in R}  
\text{Re} \Big[
\frac{1}{c} \text{Li}_2 \Big( e^{-i\big(\rho(\theta) -\alpha\big)  - 2 c j }   \Big)
\Big]^{+1/2}_{-1/2}
,
\label{dilog}
\end{align} 
where Li$_2$ is the dilogarithmic function.
In the following section, we discuss the phases of $\mathop{\rm SU}(3)$ gauge theory based on this dilogarithmic potential.
\paragraph{Contribution from the vector multiplet}
After large $\mathcal{R}$-charge procedure $\Delta \to \infty$, we have
\begin{align}
\mathcal{Z}_{\text{1-loop}}^{vec(reg)}
\to
\prod_{i > j} 
|2\sin \frac{\theta_i -\theta_j}{2}|^2.
\end{align}
This is \textit{the Haar measure}.
As commented in \cite{Cossu:2013ora},
this is not the dynamical contribution but the Jacobian caused by diagonalizing the Wilson line phase $\theta$,
and we should not take it into account.
It means we cannot break the gauge symmetry only with SYM.
This interpretation does not conflict with the known results based on the perturbative calculation \cite{Hatanaka:2011ru}.
\subsection{Notations of $\mathop{\rm SU}(3)$ phases}

We use the conventional names for particular $(\theta_1, \theta_2)$ sets \cite{Cossu:2013nla}.
We use the following names :
\begin{align}
\mathop{\rm SU}(3)\text{ symmetric configurations  } &
\left\{ \begin{array}{ll}
A_1 : (\theta_1 , \theta_2) = (0,0) \\
A_2 : (\theta_1 , \theta_2) = (\frac{2}{3} \pi , \frac{2}{3} \pi) \\
A_3 : (\theta_1 , \theta_2) = (-\frac{2}{3} \pi , -\frac{2}{3} \pi)
\end{array} \right. ,
\label{As}
\\
\mathop{\rm SU}(2) \times \mathop{\rm U}(1)\text{ symmetric configurations  } &
\left\{ \begin{array}{ll}
B_1 : (\theta_1 , \theta_2) = (0,\pi) \\
B_2 : (\theta_1 , \theta_2) = (\frac{2}{3} \pi ,- \frac{1}{3} \pi) \\
B_3 : (\theta_1 , \theta_2) = (-\frac{2}{3} \pi , \frac{1}{3} \pi)
\end{array} \right. ,
\label{Bs}
\\
\mathop{\rm U}(1) \times \mathop{\rm U}(1)\text{ symmetric configurations  } &
\left\{ \begin{array}{ll}
C : (\theta_1 , \theta_2) = (0,\frac{2}{3} \pi ) 
\end{array} \right. .
\label{Cs}
\end{align}
Figure \ref{phases} explains the positions for phases $A,B,C$, in the $(\theta_1 , \theta_2)$ plane.
The symmetries $\mathop{\rm SU}(3)$, $\mathop{\rm SU}(2)\times \mathop{\rm U}(1)$ and $\mathop{\rm U}(1)\times \mathop{\rm U}(1)$ in \eqref{As}, \eqref{Bs} and \eqref{Cs} correspond to the remaining gauge symmetry in the context of the Hosotani mechanism.

\begin{figure}[t]
  \begin{center}
   \includegraphics[width=100mm]{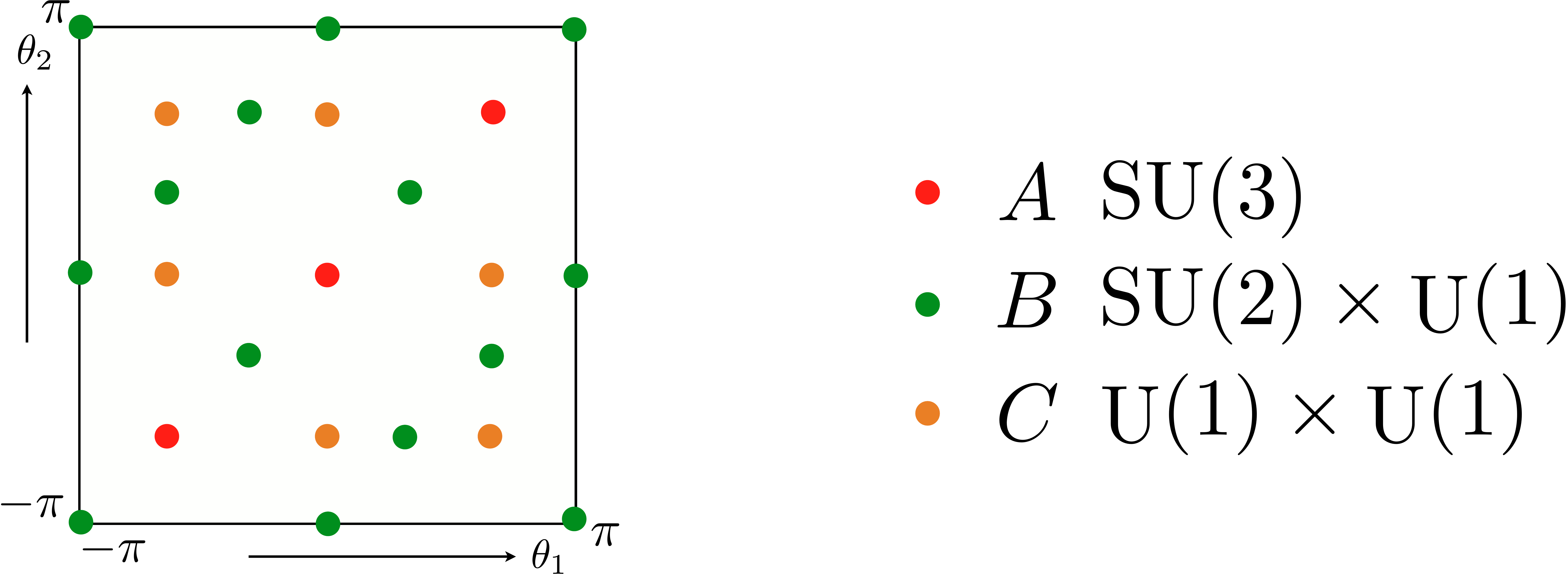}
  \end{center}
    \caption{Names for each configuration}
  \label{phases}
\end{figure}
\newcommand{\finite}{section}
\newcommand{\infinite}{next section}

\section{Finite $\Delta$ and the finite size effects}
Here, we do not intend to discuss the symmetry breaking, but \textit{the finite size effects} caused by GNO charge.
Small $l$ corresponds to the small $S^2$.
As we commented in Section 2, we consider fixed $c=\Delta\pi R/l$ \eqref{c}.
Combining small $l$ and fixed $c$, we expect that the finite size effect emerges with small $\Delta$.
For small $\Delta$, we should not use the dilogarithmic effective potential \eqref{dilog} but \eqref{eff} which depends on GNO charge.
One may wonder how we should determine the precise values of GNO charges $(m_1 , m_2)$.
However, it is clear from \eqref{eff} that the effects of GNO charge will be dropped when we take large $l$.
We assume this ambiguity for choosing $(m_1 ,m_2)$ values itself is also one of the finite size effects.
In this \finite, we observe what happens when we take $\Delta =2$ and $(m_1,m_2)=(1,1)$ as an examination of the finite size effects.
\newpage
%


\subsection{Fundamental matter}
%

 \begin{figure}[t]
 \begin{center}
   \includegraphics[width=150mm]{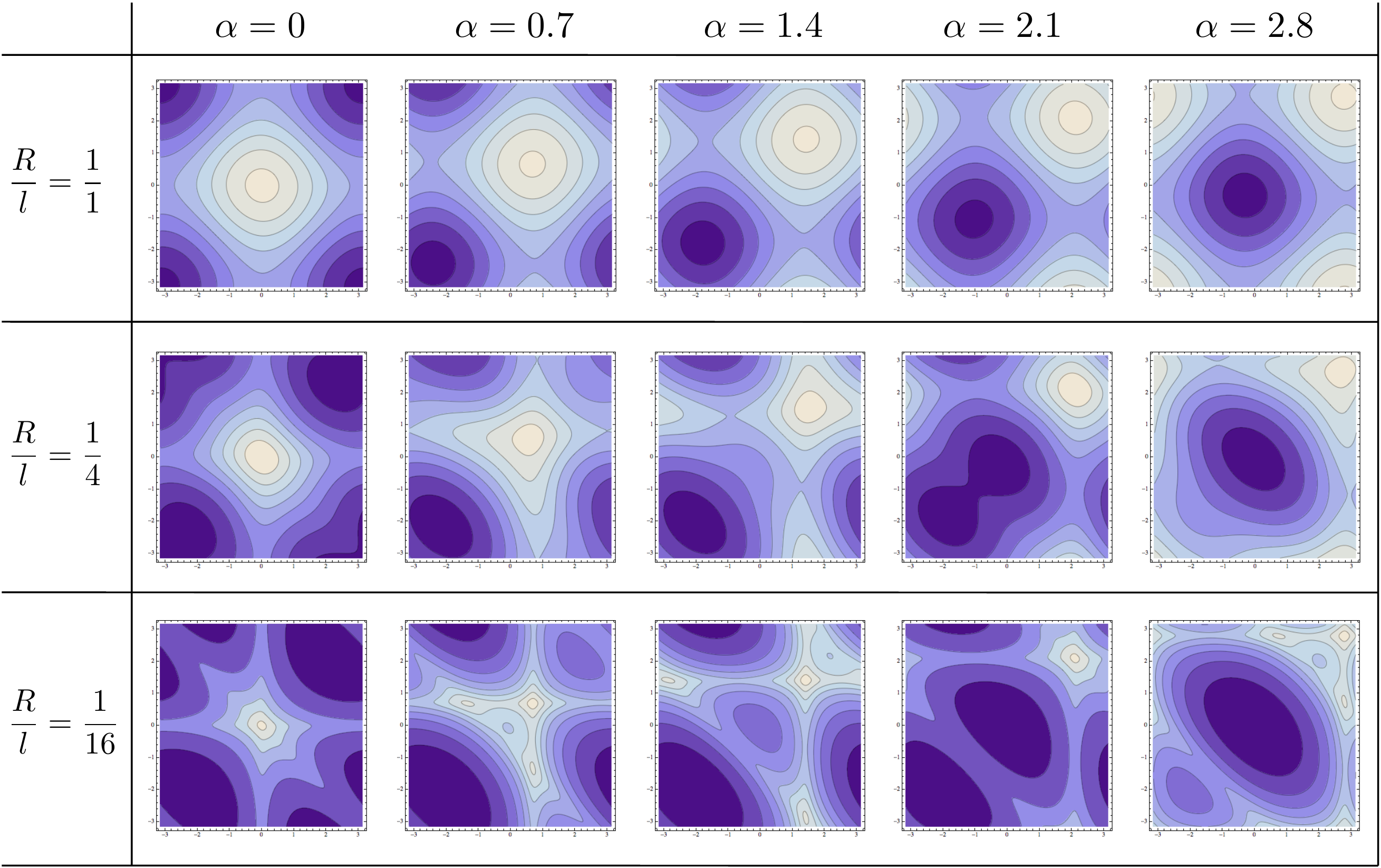}
\end{center}
   \caption{
   Contour plots of the effective potential for the fundamental matter, $\rho = fd$.
   Smaller values of the effective potential correspond to darker colors.
  The column corresponds to $R/l$ =(the size of $S^1$)/(the size of $S^2$). The row corresponds to the imaginary chemical potential $\alpha$.    \label{m1m2=11fd}}
\end{figure}

See Figure \ref{m1m2=11fd}.
We plot the effective potentials for the fundamental matter, $\rho = fd$, with various ratios $R/l$ (the column) and imaginary chemical potentials $\alpha\in [0,2.8]$ (the low).
There are two important things we shall explain.
\paragraph{Splitting locations of minima}
The first row ($R/l=1$) in Figure \ref{m1m2=11fd}  shows false minima for $(\theta_1, \theta_2)$.
For example, when the imaginary chemical potential $\alpha = 0$, $(\theta_1, \theta_2) = (0 , 0)$ looks the minimum.
However, this false vacuum is caused by the non-zero values $(m_1 ,m_2) = (1,1)$, and splits into two true vacua when we take large $l$.
It is easier to observe this splitting with $\alpha = 2.1$ column.
\paragraph{About RW transition}
Once we turn on the imaginary chemical potential $\alpha$, an interesting phenomena occur as shown in the lows of Figure \ref{m1m2=11fd}.
In the $R/l = 1/16$ low, we can see discrete change of the locations of the minima at $\alpha= 2.1$ which is known so-called RW transition \cite{Roberge:1986mm}.
On the other hand, in the $R/l = 1$ low, the minimum looks moving continuously along the line $\theta_2 = \theta_1$.
In the intermediate region \textit{i.e.} the $R/l = 1/4$ low,
we observe continuous move of the minimum in $0\leq \alpha\leq 1.4$.
However, a very quick transition of the minimum occurs around $\alpha = 2.1$.
\subsection{Adjoint matter}

 \begin{figure}[t]
 \begin{center}
   \includegraphics[width=150mm]{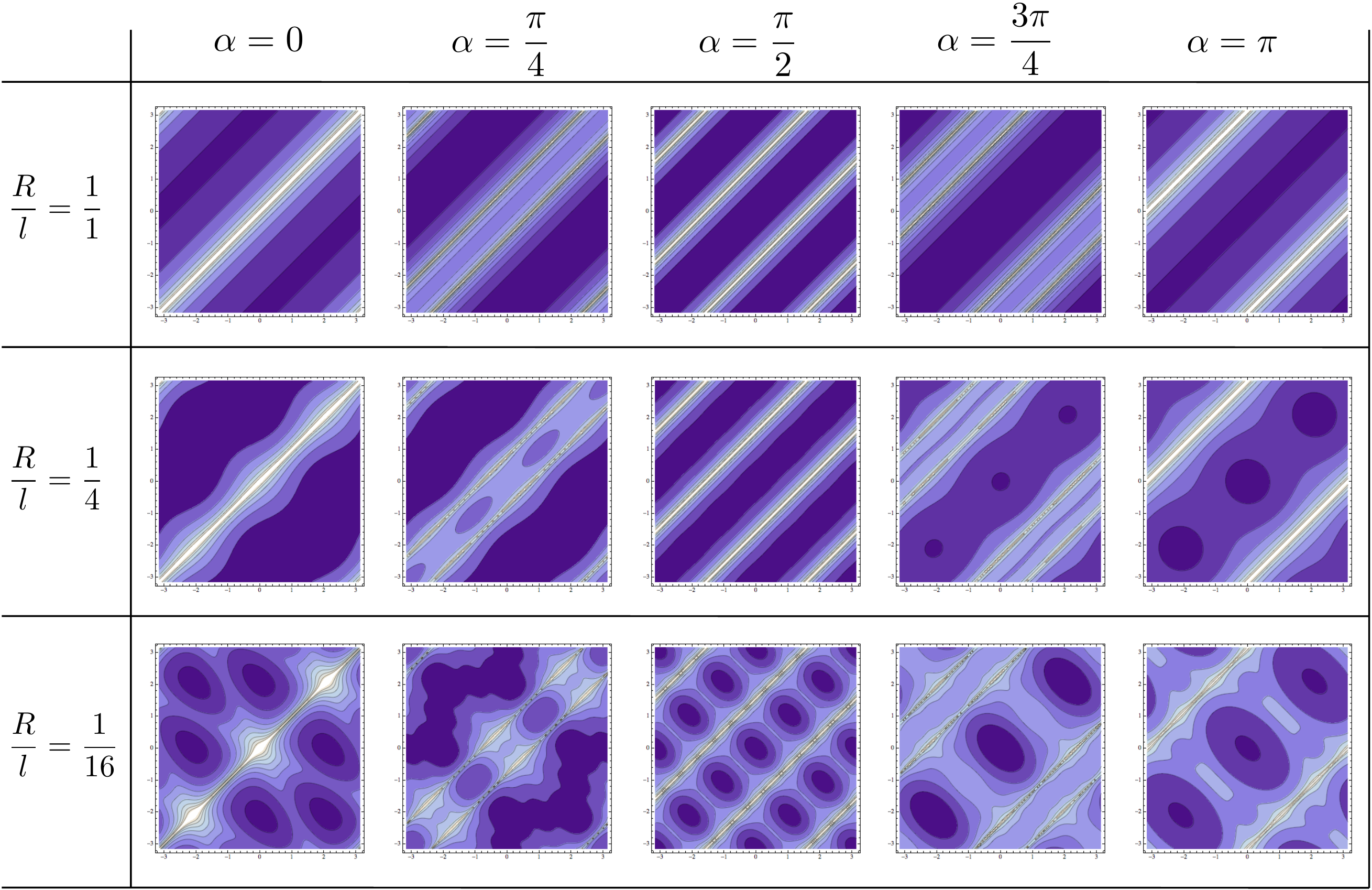}
\end{center}
   \caption{\label{m1m2=11ad}
Contour plots of the effective potential for adjoint matter, $\rho = ad$. Smaller values of the effective potential correspond to darker colors. The column corresponds to $R/l$ =(the size of $S^1$)/(the size of $S^2$). The row corresponds to the imaginary chemical potential $\alpha$.}
\end{figure}
See Figure \ref{m1m2=11ad}.
We plot the effective potentials for the adjoint matter, $\rho = ad$, with various ratios $R/l$ (the column) and imaginary chemical potentials $\alpha\in [0, \pi]$ (the low).
There are also two important things we shall explain.
\paragraph{Degenerated minima}
The first row ($R/l=1$) in Figure \ref{m1m2=11ad} shows false \textit{degenerated} minima for $(\theta_1, \theta_2)$.
Rigorously, the potential is not degenerated but has very slight depth around the minima.
The degeneracy is truly realized in $l \to +0$ limit.
Such a ill behavior of the minima is also caused by the presence of $(m_1 ,m_2)$.
In fact, such behavior vanishes as we take large $l$.
Therefore this is caused by the finite size effect.
For example, with $\alpha=\pi$, one can see that the degeneracy becomes weaker as $l$, the radius of $S^2$, becomes larger.
\paragraph{RW-like transition}
We have ``jumps'' of the location of degenerated vacua.
With $\alpha \sim 0$, the vacua around $C$, $B$ phases are preferred.
On the other hand, when we turn $\alpha \sim \pi$, the vacua around $A$ phase is preferred.
This jumping structure is observed both in the region $R \sim l$ and the region $R \gg l$.
This means such phenomena are not caused by the finite size, but come from universal structure of the $\mathop{\rm SU}(3)$ gauge theory with adjoint matters.
We will see later that this structure emerges even in large $\mathcal{R}$-charge limit.
\clearpage
\section{Large $\Delta$ and the symmetry breaking}
 \begin{figure}[t]
 \begin{center}
  \includegraphics[width=15cm]{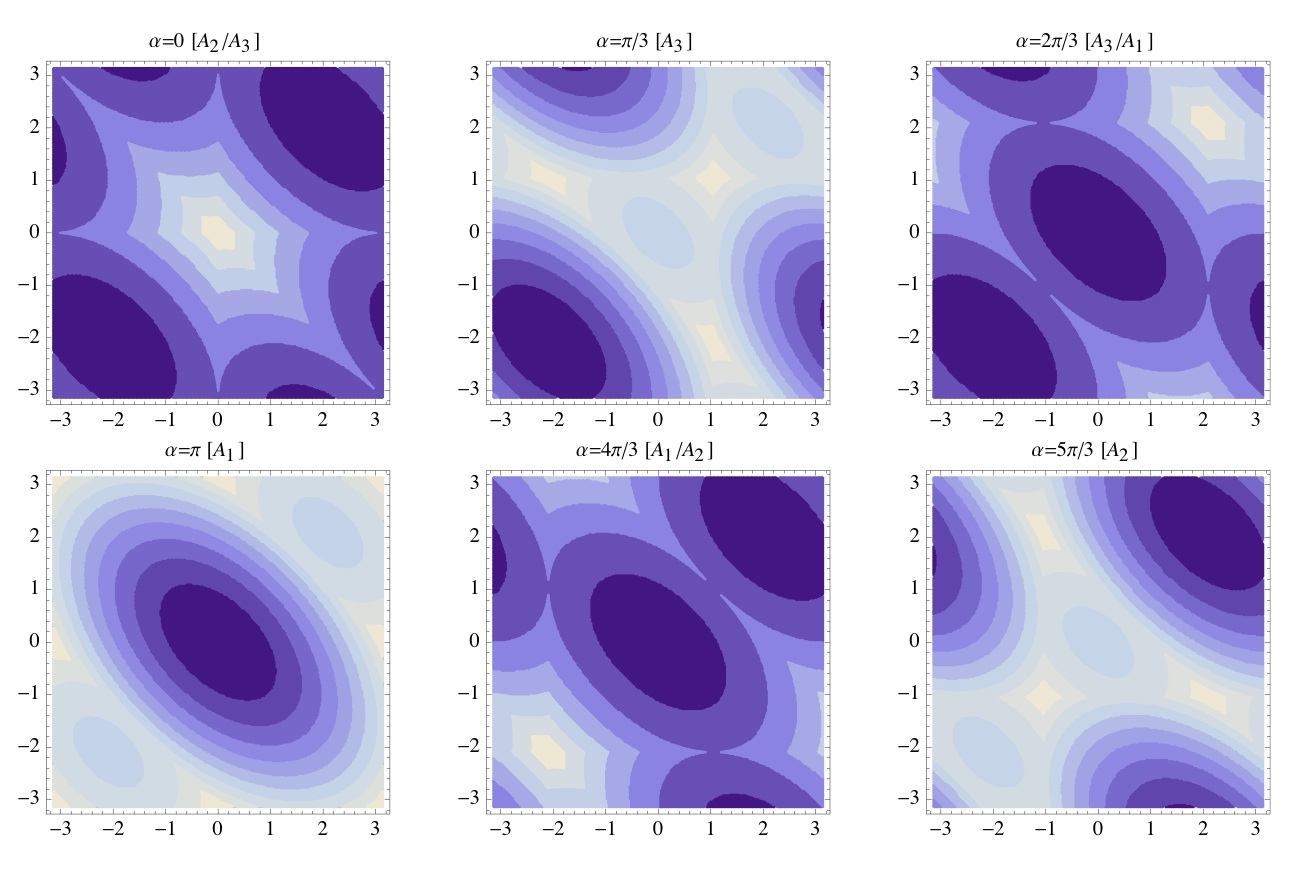}
\end{center}
   \caption{\label{figs3:fund_d10000}
   Contour plots of the effective potential for the fundamental matter, $\rho = fd$.
   Smaller values of the effective potential correspond to darker colors. 
   From top left to top right are $\alpha=0, \pi/3, 2\pi/3$ and
  bottom lines are $\alpha= \pi, 4\pi/3, 5\pi/3$.
  Vertical and horizontal axis are $\theta_1$ and $\theta_2$ in each figure.
  From left top to right bottom panels correspond to
  $A_2/A_3$, $A_3$, $A_3/A_1$, $A_1$, $A_1/A_2$ and $A_2$ phase respectively.}
\end{figure}
\paragraph{Analysis method}
In Section 2,
we obtain the effective potential \eqref{dilog}.
The effective potential has free parameter $c$.
In principle we can take an arbitrary value of $c$.
We check the $c$ dependence of our effective potential.
As a result, locations of minima of our effective potential do not change qualitatively for $c =50$, $500$, $5000$.
In this section, we show contour plots of the effective potential with $c=5000$ (Figure \ref{figs3:fund_d10000}, Figure \ref{figs3:adj_d10000}).
\par
By the way, there exists another non-perturbative result for the phase structure based on the lattice gauge theory \cite{Cossu:2013ora}.
They measured the Polyakov loop and reconstruct the effective potential from configurations of the Polyakov loop.
In our method we can see non-perturbative exact effective potential directly.
We do not see the Polyakov loop nor other physical quantities.
\subsection{Fundamental matter}
First, we investigate $\alpha$ dependence of minima for the effective potential with the fundamental matter (Figure \ref{figs3:fund_d10000}).
This is one of our main results.
Darker regions correspond to deeper regions of our effective potential.
As we said before, a minimum of the effective potential is selected in large $\mathcal{R}$-charge limit.
Phases appear in the order of $A_3$, $A_1$, $A_2$ as $\alpha$ moves from $\pi/3$ to $5\pi/3$ with discrete transition.
This means we obtain RW transition in exact way.
\par
Compared with the perturbative result for non-supersymmetric 4 dimensional theory in \cite{Cossu:2013ora}, appearing order of $A_{1,2,3}$ phases is inverted.
This is because we use the opposite sign convention of the gauge coupling through the covariant derivative \eqref{covd}.
In this sense, our result is similar to their result.
However our effective potential does not depend on any coupling constant.
This fact indicates there is no difference between the strong coupling limit and the weak coupling limit.
We comment on this issue later.\par
\begin{figure}[t]
 \begin{center}
  \includegraphics[width=15cm]{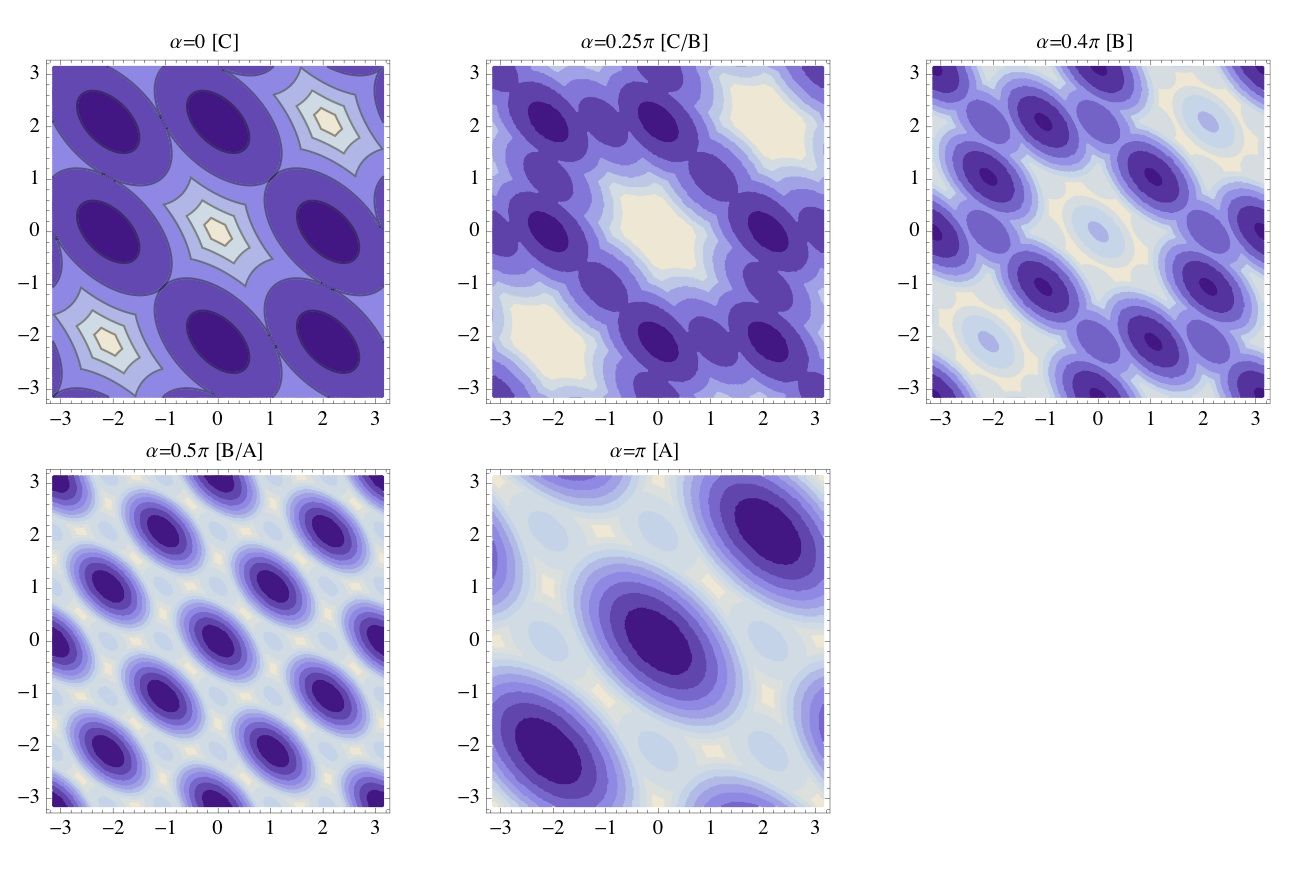}
\end{center}
   \caption{
    \label{figs3:adj_d10000}
    Contour plots of the effective potential for adjoint matter, $\rho = ad$.
    Smaller values of the effective potential correspond to darker colors.
   From top left to top right are $\alpha=0, 0.25\pi, 0.4\pi$ and
  bottom lines are $\alpha= 0.5\pi$ and $\pi$.
  Vertical and horizontal axis are $\theta$.
  From left top to right bottom panels correspond to
  $C$, $C/B$, $B$, $B/A$ and $A$ phase respectively.}
\end{figure}

\subsection{Adjoint matter}
Next, we investigate $\alpha$ dependence of minima for the effective potential with the adjoint matter.
Figure \ref{figs3:adj_d10000} is our second main result.
These contour plots show discrete phase transitions.
Phases appear in the order of $C$, $B$, $A$ as $\alpha$ moves from $0$ to $\pi$.
When we increase $\alpha$ from $\pi$ to $2\pi$, the phases move to $A$, $B$ and go back to $C$.\par
Again we get the similar result via non-perturbative method.
Compared with the perturbative result for non-supersymmetric 4 dimensional theory in \cite{Cossu:2013ora}, they have shown critical points for the boundary condition.
The global minima of the effective potential with $c=0$ are located at
\begin{align}
A_{1,2,3}\quad &\text{for}\quad 0.5 \pi \leq \alpha \leq \pi,\nonumber\\
B_{1,2,3}\quad &\text{for}\quad 0.3 \pi \leq \alpha \leq 0.5 \pi,\nonumber\\
C\quad &\text{for}\quad \alpha \leq 0.3 \pi.
\end{align}
We determine the critical values for the imaginary chemical potential $\alpha=0.3 \pi$, $0.5\pi$ in numerical calculation.
We expect that these values are determined analytically for $c\neq0$.
\section{Conclusion and Discussion}

\paragraph{Summary and Conclusion}
 \begin{figure}[t]
 \begin{center}
  \includegraphics[width=14cm]{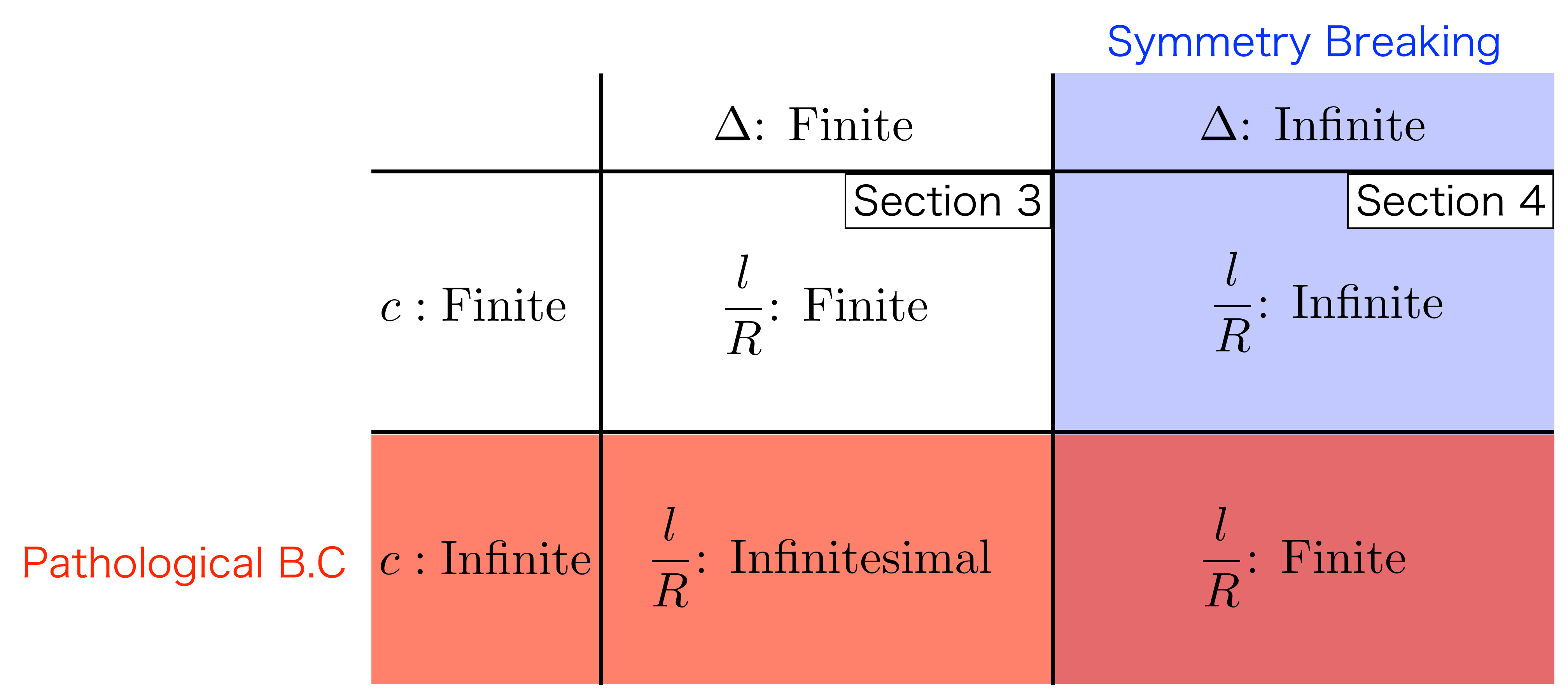}
\end{center}
   \caption{$l$ is the radius of $S^2$. $R$ is the radius of $S^1$. Red colored region defines the pathological boundary conditions with matters, and we do not consider in this paper.
In blue colored region, the symmetry breaks due to large $\Delta$.
\label{table}
}
\end{figure}
We calculated the effective potential of the Wilson line phase for SYM theory with two matters on $S^2\times S^1$ via the localization technique.
See Figure \ref{table}.
Red colored region defines the pathological boundary conditions with matters, and we do not consider in this paper.
In blue colored region, the symmetry breaks due to large $\Delta$.
Our main target region is the one with finite $c$ and infinite $\Delta$. In this region, the volume is also infinite because $l$ is infinite. So as a result, we have not only the large $\mathcal{R}$-charge limit but also the thermodynamic limit as noted in Introduction.

We checked the finite size effects for our effective potential in Section 3.
For finite $\Delta$, the effective potential \eqref{eff} is affected by the existence of the GNO monopole on $S^2$.
The GNO monopole exists only on $S^2$, such a deformation of the effective potential is naturally understood as a finite size effect.
This is because the effect disappears by taking large $l$ limit.\par
On the other hand, the effective potential is written by the dilogarithmic function \eqref{dilog} for large $\Delta$.
We investigated this effective potential for matter fields in fundamental representation
and adjoint representation at large $\mathcal{R}$-charge limit.
We found the phase transition had occurred in this SUSY gauge theory non-perturbatively both with
the fundamental matter and the adjoint matter.
In general, this phenomenon for the fundamental matter is called RW transition \cite{Roberge:1986mm}.
This fact supports our analysis.

In the fundamental matter case, phases appear in the order of $A_3$, $A_1$, $A_2$ as $\alpha$ moves from $\pi/3$ to $5\pi/3$.
The critical $\alpha$s coincide with the ones in the perturbative one-loop effective potential for non-supersymmetric theory on $R^3\times S^1$ \cite{Cossu:2013ora}.

Phases with the adjoint matter appear in the order of $C$, $B$, $A$ as $\alpha$ moves from $0$ to $\pi$.
When we increase $\alpha$ from $\pi$ to $2\pi$, the phases move to $A$, $B$ and go back to $C$.
The critical values for the imaginary chemical potential $\alpha=0.3 \pi$, $0.5\pi$ are determined by numerical calculation.
\paragraph{Discussion}
We have 2 open questions for this model.
First, in usual flat supersymmetric theories,
contributions from bosons and fermions are canceled out completely.
Therefore this effective potential for such models becomes flat(trivial).
However, in our case, the potential is nontrivial.
We guess this kind of a phenomenon caused by non-zero curvature effect.
We need to investigate why the potential becomes nontrivial.
Second, several previous works for RW transition had coupling constant dependence.
However our model has no coupling constant after using the localization technique.
In other words, our effective potential has no sensitivity for coupling constant.
It is interesting to use other localization results which do depend on coupling constants. 

\paragraph{Some issues}
We would like to point out 2 issues of our analysis.
First,
we \textit{assume} infinite volume limit is not spoiled by the GNO monopoles. 
In other words, we just dropped the contribution of $m$ when we take large $\mathcal{R}$-charge limit.
This assumption may be problematic because there are always monopoles with arbitrary large $|m|$, and in this case, the dropping of $m$ becomes subtle.
The second issue is related to the large $\mathcal{R}$-charge limit itself.
In order to cause the symmetry breaking, we argue that the large $\Delta$ is necessary.
However, $\Delta$ looks bounded in certain region.
This restriction comes, naively speaking, from the sign of the quadratic potential for $\phi$ in the matter Lagrangian \eqref{Smat}.
If one wants to overcome this undesirable situation, it may be possible to recover it by adding certain SUSY-exact terms.
Another way to recover it is 
taking $c=0$.
In this case, we have $\Delta  \ll l$ and this condition makes the quadratic potential for $\phi$ to be zero.
However these remedies are somewhat subtle.
And these problems look very crucial.
So we have to find better solutions.

\paragraph{Future direction}
There are some extensions.
One direction is to change background geometries.
For example there are localization calculation results on $D^2$ \cite{Sugishita:2013jca, Hori:2013ika, Honda:2013uca} and $D^2\times S^1$ \cite{Sugishita:2013jca}.
As more phenomenological setup, we should consider theory on $M \times S^1/Z_2$ or Randall-Sundram spacetime.
The Wilson line phase comes from these $S^1$ and $S^1/Z_2$.
Another direction is the localization method in higher dimensional theories.
For instance, we could apply results on $S^3 \times S^1$ \cite{Romelsberger:2005eg} and $CP^2 \times S^1$ \cite{Kim:2013nva} to the exact calculation of the effective potential for the Wilson line phase.

\section*{Acknowledgment}
We inspired at workshop ``Extra Dimension 2013" at Osaka University.
We appreciate the discussion with Hidenori Fukaya, Hisaki Hatanaka, Yutaka Hosotani and Satoshi Yamaguchi.
We would like to thank  Koji Hashimoto, Kin-ya Oda, Tetsuya Onogi, Toshihiro Matsuo and our colleagues for encouragements to write this paper.
This work was supported in part by JSPS KAKENHI grants, No. 13J01891(A.Tanaka) and No. 13J01861(T.S.).
\appendix
\section{SUSY on $S^2 \times S^1$ and the exact results}
\paragraph{Killing spinors}
As one can find in \cite{Gang:2009wy, Imamura:2011su, TanakaMori}, it is sufficient for defining supersymmetric field theories on $S^2 \times S^1$ to find so-called Killing spinors\footnote{See \cite{Closset:2012ru} for more systematic approach.}.
We take the following 2 Killing spinors:
\begin{align}
\e
=
e^{\frac{1}{2} (\frac{y}{l} + i \varphi)}
\begin{pmatrix}
\cos \frac{\vartheta}{2} \\
\sin \frac{\vartheta}{2}
\end{pmatrix}
,
\qquad
\oep
=
e^{\frac{-1}{2} (\frac{y}{l} + i \varphi)}
\begin{pmatrix}
\sin \frac{\vartheta}{2} \\
\cos \frac{\vartheta}{2}
\end{pmatrix}.
\label{KSEf}
\end{align}
These spinors satisfy the following equations,
\begin{align}
\D_\mu \e = \frac{1}{2l} \g_\mu \g_3 \e,
\qquad
\D_\mu \oep = \frac{-1}{2l} \g_\mu \g_3 \oep,
\label{KSE}
\end{align}
where we take vielbein as
\begin{align}
e^1 = l d \vartheta,
\quad
e^2 = l \sin \vartheta d \varphi,
\quad
e^3 = d y.
\end{align}
\paragraph{Vector multiplet}
We can construct the $\mathcal{N}=2$ vector multiplet $V=(A_\mu , \sigma , \bar{\lambda}, \lambda, D)$ on $S^2 \times S^1$ by using $\e, \oep$ defined in \eqref{KSEf}.
We use SUSY transformation defined in \cite{Hama:2010av}.
Though their manifold is $S^3$, their SUSY construction is enough generic to use even on $S^2 \times S^1$.
However, one cannot define SUSY invariant theory not only with the SUSY transformations, but also the $S^1$ boundary conditions for the component fields:
\begin{align}
&A_\mu (\vartheta,\varphi,y+2\pi R)
=
A_\mu (\vartheta,\varphi,y),
\label{vbc1}
\\
&\si (\vartheta,\varphi,y+2\pi R)
=
\si (\vartheta,\varphi,y),
\label{vbc2}
\\
&\lam (\vartheta,\varphi,y+2\pi R)
=
e^{\frac{\pi R}{l}} \lam (\vartheta,\varphi,y),
\label{vbc3}
\\
&\olam (\vartheta,\varphi,y+2\pi R)
=
e^{-\frac{\pi R}{l}} \olam (\vartheta,\varphi,y),
\label{vbc4}
\\
&D(\vartheta,\varphi,y+2\pi R)
=
D (\vartheta,\varphi,y).
\label{vbc5}
\end{align}
Note that the $\lam ,\olam$ have nontrivial scaling once they wrap the $S^1$.
This scaling boundary condition comes from the $y$ dependence of $\e, \oep$ in \eqref{KSEf}.
One can guess that only $\mathcal{R}$-charged fields have the scaling.
In fact, it becomes clear once we write down the definition of the \textit{index} \cite{Kim:2009wb, Gang:2009wy, Imamura:2011su, TanakaMori}.
Within these component fields and the Lagrangian \eqref{SYMl}, one can derive the following result \cite{Kim:2009wb, Gang:2009wy, Imamura:2011su},
\begin{align}
\int \mathcal{D} V
e^{-  \frac{1}{g_{\text{YM}}^2} \int \sqrt{g} \mathcal{L}_{\text{SYM}}}
=
\sum _{m \in \mathbb{Z}}
\frac{1}{\text{sym}}
\int_{- \pi}^{+\pi} \prod \frac{d \theta_i}{2 \pi R}
\mathcal{Z}_{1-loop}^{(vec)},
\label{vecind}
\end{align}
where
\begin{align}
\mathcal{Z}_{1-loop}^{(vec)}
=
\prod_{i \neq j} 
\prod_{n=-\infty}^{\infty}
\prod_{J = 0}^{\infty}
\frac{J + |\frac{m_i -m_j}{2}| +i \big(\frac{l }{ R} n - \frac{l }{ R}\frac{\theta_i - \theta_j}{2 \pi} \big)  }{J+1 + |\frac{m_i - m_j}{2}|-i \big(\frac{l }{ R} n - \frac{l }{ R}\frac{\theta_i - \theta_j}{2 \pi} \big) }
.
\label{vecl1}
\end{align}
Note that the result \eqref{vecl1} does not depend on the coupling constant $g_{\text{YM}}$.
This is the consequence caused by a fact, the Lagrangian \eqref{SYMl} is \textit{SUSY-exact}.
Here, $n$ represents the Kaluza-Klein mode and $J$ corresponds to the angular momentum with respect to the $S^2$.
The meaning of $\theta_i$, $m_i$ is explained in Section 2.
We can simplify \eqref{vecl1} by using the symmetry $n \to -n$, and $\theta_i \leftrightarrow  \theta_j$ as follows
\begin{align}
\eqref{vecl1}
&=
\prod_{i \neq j} 
\prod_{n=-\infty}^{\infty}
\Big( 0 + |\frac{m_i -m_j}{2}| +i \big(\frac{l }{ R} n - \frac{l }{ R}\frac{\theta_i - \theta_j}{2 \pi} \big)   \Big)
\notag \\
&\xrightarrow{\zeta\text{-reg}}
\prod_{i > j} 
\Big| 2
\sin
\Big( 
\frac{\theta_i - \theta_j}{2} + i \frac{\pi R}{l} \frac{m_i - m_j}{2}
 \Big)
 \Big|^2
 , \label{vecl2}
\end{align}
where we use the zeta function regularization in the final step.
\paragraph{Matter multiplet}
We can also define the matter multiplets $\Phi=
 (\phi , \psi , F), \overline{\Phi}=
 (\ophi , \opsi , \oF)$ which couple with the vector multiplet via the gauge symmetry \cite{Hama:2010av}.
 As well known, we can assign arbitrary $\mathcal{R}$-charge $\Delta_\Phi$ with $\Phi$.
 We have to tune the $S^1$ boundary conditions for the component fields as follows,
\begin{align}
&\phi (\vartheta,\varphi,y+2\pi R)
=
e^{- \Delta_\Phi \frac{\pi R}{l} +\mu} \phi (\vartheta,\varphi,y),
\label{mbc1}
\\
&\psi (\vartheta,\varphi,y+2\pi R)
=
e^{- (\Delta_\Phi-1) \frac{\pi R}{l} + \mu}\psi (\vartheta,\varphi,y),
\label{mbc2}
\\
&F (\vartheta,\varphi,y+2\pi R)
=
e^{- (\Delta_\Phi-2) \frac{\pi R}{l} + \mu} F (\vartheta,\varphi,y),
\label{mbc3}
\\
&\ophi (\vartheta,\varphi,y+2\pi R)
=
e^{ \Delta_\Phi \frac{\pi R}{l} - \mu} \ophi (\vartheta,\varphi,y),
\label{mbc4}
\\
&\opsi (\vartheta,\varphi,y+2\pi R)
=
e^{ (\Delta_\Phi-1) \frac{\pi R}{l} - \mu}\opsi (\vartheta,\varphi,y),
\label{mbc5}
\\
&\oF (\vartheta,\varphi,y+2\pi R)
=
e^{ (\Delta_\Phi-2) \frac{\pi R}{l} - \mu} \oF (\vartheta,\varphi,y),
\label{mbc6}
\end{align}
in order to preserve supersymmetry.
Through the well known argument, we have no degenerate vacua with respect to $\Phi$ with the Lagrangian \eqref{Smat}.
Therefore, the only nontrivial contribution comes from by inserting the following function into \eqref{vecind},
\begin{align}
\mathcal{Z}_{\text{1-loop}} ^{mat}
&=
\prod_{\rho \in R} 
\prod_{n=-\infty}^{\infty}
\prod_{J = 0}^{\infty}
\frac{J +1- i\big(\frac{ l}{ R}n - \frac{l }{2 \pi R} ( \rho(\theta) +  i \mu) \big) -\frac{\Delta_\Phi}{2}  +|\frac{\rho(m)}{2}|   }{J + i\big(\frac{ l}{ R}n - \frac{l }{2 \pi R} ( \rho(\theta) +  i \mu) \big) + \frac{\Delta_\Phi}{2}  +|\frac{\rho(m)}{2}|   }
.
\label{matl1}
\end{align}
There is no coupling constant as same as the case of the vector multiplet, and it comes from the SUSY-exactness of the matter Lagrangian \eqref{Smat}.
Unfortunately, one cannot simplify it as we do in \eqref{vecl2}.
In order to overcome this situation, we consider not only one $\Phi$, but two matter multiplets $\Phi_1 ,\Phi_2$ as we explained in Section 2.
In this case, we can simplify the results as
\begin{align}
&
\mathcal{Z}_{\text{1-loop}} ^{mat1,2}
=\mathcal{Z}_{\text{1-loop}} ^{mat1}
\times
\mathcal{Z}_{\text{1-loop}} ^{mat2}
\notag \\
&=
\prod_{\rho \in R} 
\prod_{n=-\infty}^{\infty}
\prod_{J = 0}^{\infty}
\Big(
\frac{J +1- i\big(\frac{ ln}{ R} - \frac{l ( \rho(\theta) - \alpha)}{2 \pi R} \big) -\frac{\Delta}{2}  +|\frac{\rho(m)}{2}|   }{J + i\big(\frac{ ln}{ R} - \frac{l ( \rho(\theta) - \alpha)}{2 \pi R}  \big) + \frac{\Delta}{2}  +|\frac{\rho(m)}{2}|   }
\Big) \Big(
\frac{J +1- i\big(\frac{ ln}{ R} + \frac{l ( \rho(\theta) - \alpha)}{2 \pi R}  \big) -\frac{\Delta}{2}  +|\frac{\rho(m)}{2}|   }{J + i\big(\frac{ ln}{ R} + \frac{l ( \rho(\theta) -\alpha)}{2 \pi R}  \big) + \frac{\Delta}{2}  +|\frac{\rho(m)}{2}|   }
\Big)
\notag \\
&\xrightarrow{\zeta\text{-reg}}
\left\{ \begin{array}{ll}
\prod_{\rho \in R} 
\prod_{J = 1- \frac{\Delta}{2}}^{\frac{\Delta}{2}-1}
\Big| 2
\sin 
\Big( \frac{(\rho(\theta) -\alpha)}{2} +i \frac{\pi R}{l} \big(  |\frac{ \rho(m)}{2} | + J \big)
\Big) 
\Big|^2
& \quad ( \Delta -1 \in \mathbb{N} )  \\
1  & \quad (\Delta =1) \\
\prod_{\rho \in R} 
\prod_{J =  \frac{\Delta}{2}}^{-\frac{\Delta}{2}}
\Big| 2
\sin 
\Big( \frac{(\rho(\theta) -\alpha)}{2} +i \frac{\pi R}{l} \big(  |\frac{ \rho(m)}{2} | + J \big)
\Big) 
\Big|^{-2} & \quad (1- \Delta  \in \mathbb{N} )
\end{array} \right.
.\label{matl3}
\end{align}
We take $\Delta-1\in \mathbb{N} $ throughout this paper.
\providecommand{\href}[2]{#2}\begingroup\raggedright\endgroup

\end{document}